\documentclass[sigconf]{acmart}

\usepackage{natbib}
\usepackage{booktabs}
\usepackage{multicol}
\usepackage{subcaption}
\usepackage{colortbl}
\usepackage{longtable}
\usepackage{lipsum}
\usepackage{svg}
\usepackage[ruled,linesnumbered]{algorithm2e}
\usepackage{tabularx}
\usepackage[shortlabels]{enumitem}
\usepackage{textcomp}

\newcolumntype{L}{>{\raggedright\arraybackslash}X}

\usepackage{changepage}
\usepackage{framed}

\newcommand{\sbar}[1]{{\color{darkgray}\rule{\dimexpr 1cm * #1 / 100}{5pt}\color{lightgray}\rule{\dimexpr 1cm * (100 - #1) / 100}{5pt}}}

\newcommand{\bi}{\begin{itemize}}
\newcommand{\ei}{\end{itemize}}

\makeatletter
 {\par\unskip\endMakeFramed}
\makeatother

\definecolor{formalshade}{rgb}{0.93,0.93,0.93}
\newcommand{\gray}{\cellcolor{lightgray}}

\definecolor{darkblue}{rgb}{0.2, 0.2, 0.2}

\newenvironment{formal}{%
  \def\FrameCommand{%
    \hspace{1pt}%
    {\color{darkblue}\vrule width 2pt}%
    {\color{formalshade}\vrule width 4pt}%
    \colorbox{formalshade}%
  }%
  \MakeFramed{\advance\hsize-\width\FrameRestore}%
  \noindent\hspace{-1pt}
  \begin{adjustwidth}{}{7pt}%
  \vspace{2pt}\vspace{2pt}%
}
{%
  \vspace{3pt}\end{adjustwidth}\endMakeFramed%
}

\newcommand{\BLUE}{\color{black}}
\newcommand{\BLACK}{\color{black}}

\AtBeginDocument{%
  \providecommand\BibTeX{{%
    \normalfont B\kern-0.5em{\scshape i\kern-0.25em b}\kern-0.8em\TeX}}}

\setcopyright{acmcopyright}

\copyrightyear{2022}
\acmYear{2022}
\setcopyright{acmcopyright}\acmConference[MSR '22]{19th International Conference on Mining Software Repositories}{May 23--24, 2022}{Pittsburgh, PA, USA}
\acmBooktitle{19th International Conference on Mining Software Repositories (MSR '22), May 23--24, 2022, Pittsburgh, PA, USA}
\acmPrice{15.00}
\acmDOI{10.1145/3524842.3528458}
\acmISBN{978-1-4503-9303-4/22/05}

\acmConference[MSR 2022]{MSR '22: Proceedings of the 19th International Conference on Mining Software Repositories}{May 23–24, 2022}{Pittsburgh, PA, USA}
\acmPrice{15.00}
\acmISBN{978-1-4503-XXXX-X/18/06}

\begin{document}

\title{How to Improve Deep Learning for Software Analytics\\(a case study with   code smell detection)}

\author{Rahul Yedida}
\orcid{0000-0003-2069-5949}
\affiliation{%
  \institution{Dept. of Computer Science, NC State University, USA}
  \country{}
}
\email{ryedida@ncsu.edu}

\author{Tim Menzies, \textit{IEEE Fellow}}
\affiliation{%
  \institution{Dept. of Computer Science, NC State University, USA}
  \country{}
}
\email{timm@ieee.org}

\begin{abstract}
To reduce technical debt and make code more maintainable, it is important to be able to warn programmers about code smells. State-of-the-art code small detectors use deep learners, usually without exploring alternatives. For example, one promising  alternative is GHOST (from TSE'21)
that relies  on a combination of hyper-parameter optimization of feedforward neural networks and a novel oversampling technique.

The prior study from TSE'21  proposing this novel ``fuzzy sampling'' was somewhat limited in that the method was tested on defect prediction, but nothing else. Like defect prediction, code smell detection datasets have a class imbalance (which motivated ``fuzzy sampling"). Hence, in this work we test if fuzzy sampling is useful for code smell detection. 
 
The results of this paper show that we can  achieve better than state-of-the-art results on code smell detection with fuzzy oversampling. For example, for ``feature envy'', we were able to achieve 99+\% AUC across all our datasets, and on 8/10 datasets for ``misplaced class''. While our specific results refer to code smell detection, they do suggest other lessons for other kinds of analytics. For example: (a) try better preprocessing before trying complex learners (b) include simpler learners as a baseline in software analytics (c) try ``fuzzy sampling'' as one such baseline. 

In order to support others trying to reproduce/extend/refute this work, all our code and data is available online at \url{https://github.com/yrahul3910/code-smell-detection}.
\end{abstract}




 \maketitle

\section{Introduction}

In their enthusiasm to try the latest and greatest method,
are researchers not  reflecting on how to best use those methods? 
For example, a  common claim is that DL supports a kind of 
automated feature engineering~\cite{zeiler2014visualizing,panda2016unsupervised,nair2010rectified,suk2014hierarchical,yamashita2018convolutional} 
that lets  data scientists   avoid
tedious manual feature engineering,  prior to running their learners.
Are these learners being applied in the best way, for software analytics?

Perhaps not.   In a recent TSE'21 paper, 
Yedida \& Menzies \cite{yedida2021value} found that, for defect prediction, they  needed to
significantly augment that automated feature engineering with a technique they called ``fuzzy sampling''. With that addition, and after comparing
to recent results
applying deep learning to software engineering, they could achieve a new state-of-the-art result.

While an interesting study, Yedida \& Menzies never tested their methods on anything else other than defect prediction. Accordingly, in this paper, we test
if their extension to deep learning helps another SE domain.
{\em
Code smell detection} is an integral part of software maintenance, and facilitates refactoring for better code quality and lesser technical debt. For example, \citet{moser2007case} provide industry evidence that refactoring, whether in general or for code smells, improves development productivity. \citet{zazworka2011investigating} showed that God Classes, one type of code smell studied in this paper, are more defect-prone. \citet{deligiannis2004controlled} showed that a design without God Classes will result in better correctness and consistency. Therefore, there is industry motivation for better code smell detectors. Moreover, like the defect prediction studied by \citet{yedida2021value}, code smell detectors also suffers from the class imbalance problem, which motivated them to develop their fuzzy sampling approach.


To satisfy that need, we start by performing a  literature review of the application of deep learning in software analytics tasks (as opposed to just code smell detection) and reporting the results. Rather than just check for feedforward networks (the basic form of deep learners, introduced in \S\ref{sec:feedforward})--which are often combined with other approaches \cite{zhuang2021software, gao2021automating,choetkiertikul2021automatically,li2020tagdc}--we search more broadly
to find a \textit{range} of deep learning methods used  in SE, and report the distribution of their use across different tasks.  Overall, we are interested in (a) what architectures are frequently used in software engineering (b) where feedforward networks stand among the choices (c) how prevalent deep learning is, in the first place. We will find that feedforward networks are significantly under-studied in SE.

Having established this, we proceed to look at the few feedforward networks that are in use, and see what distinguishes them.  We then asked if that kind of reasoning
can be applied in other SE tasks as well.

Our investigation covers the following research questions:

\underline{\textbf{RQ1:}} \textit{Can  fuzzy over-sampling achieve state-of-the-art results in code smell detection?}

To answer this, we detail our datasets, show that using the oversampling methods from   fuzzy oversampling is necessary, and then use statistical tests to check that 
fuzzy-oversampling outperforms the prior state-of-the-art. Our conclusion for this research question will be:

\begin{formal}
    \noindent
    Feed forward networks, augmented with fuzzy over-sampling,  achieves state-of-the-art results in code smell detection.
\end{formal}

Next, we will work towards a more open research question. Internally, 
this feed forward architecture
is a a very simple
neural net architecture that is now decades old.
This architecture is much simpler than the  deep learning methods that feature prominently in the current literature. Hence we ask:

\underline{\textbf{RQ2:}} \textit{Why do feedforward networks work so well?}

This is a more general question that ponders why fuzzy oversampling  works so well across multiple software analytics domains. To do so, we revisit AI literature. Specifically, \citet{hornik1989multilayer} lay out a ``universal approximation theorem'' that states that feedforward networks can model an arbitrary decision boundary. We combine that with more modern deep learning theory on feedforward networks \cite{jacot2018neural,montufar2014number}. Our conclusion from this investigation will be:

\begin{formal}
    \noindent
    With the right set of hyper-parameters, feedforward networks are ``universal approximators'';
    i.e. are theoretically applicable to many domains
\end{formal}
This obseration,
we suggestion, explains how older neural net technology defeats a more recent deep learner.

The rest of this paper is structured as follows. Section \ref{sec:background} provides a background on code smell detection, deep learning, and how deep learning has been used in software engineering. Section \ref{sec:method} discusses our method in detail. In Section \ref{sec:results}, we answer the research questions we put forward. Section \ref{sec:broader} discusses broader implications of this work. Section \ref{sec:conclusion} concludes this paper.

Before beginning, we pause to make the point that we are {\bf not} saying that older methods 
{\bf always} defeat modern deep learning. Instead we offer a  case study where a decades  old kind of neural network  defeats a new-style deep learner. We say this older method worked for fundamental reasons:
\bi
\item
When learners are tunable, they are improvable; 
\item
Older methods run very fast and so are easier to tune.
\ei
Having read the   literature on DL we can assert that this kind of comparison (of new  technology to older neural net methods) is done very rarely. In fact, in our reading of the  literature,we can find only two examples where such a comparison has been conducted and {\bf none of those two examples comes from the SE literature}. Hence we think it is a valid concern to raise at this time: can we simplify much of the neural net research in SE? 


\section{Background}
\label{sec:background}

This section offers note on our domain of study (code smell detection) and our learners (feed forward networks and deep learners).

\subsection{Code smell detection}
\label{sec:smells}

\citet{beck1999bad} proposed the idea of \textit{code smells}, which are ``certain structures in the code that suggest (sometimes they scream for) the possibility of refactoring''. They introduce 22 different code smells, of which we study four:

\begin{itemize}
    \item \textbf{Feature envy} refers to a condition when a method accesses the data of another object more than its own. This is a sign that the method should be a part of the other class instead.
    \item \textbf{Large class} refers to when a class tries to do too much. This can be identified by having too many instance variables, which may lead to duplicated code.
    \item \textbf{Long method} refers to methods that are too long and cause difficulty in comprehending their functionality and scope. \citet{beck1999bad} state that shorter methods facilitate ``explanation, sharing, and choosing'', and suggest being aggressive about decomposing methods.
    \item \textbf{Misplaced class} occurs when classes are improperly distributed, and should be moved to the correct package.
\end{itemize}

\BLUE
We use the same four code smells as prior work \cite{liu2019deep} to make a fair comparison. \BLACK Detecting code smells using machine learning is widely studied in the SE literature. \citet{azeem2019machine} present a systematic literature review of the field, but we discuss some here. \citet{schumacher2010building} study the god class smell in a commercial environment, and the efficacy of automated metric-based systems to detect it. Contrary to other researchers' beliefs, \citet{yamashita2013extent} conclude that code smells have a minor impact on software maintainability. \citet{fontana2013code} study four code smells: god class, data class, long method, and feature envy using twelve subjects from Qualitas Corpus of \citet{tempero2010qualitas}. \citet{sahin2014code} use bilevel optimization to study seven code smells over nine open-source projects. \citet{palomba2015textual} study long method detection on three software systems. \citet{fontana2016comparing} conclude that machine learning can effectively be applied for code smell detection, achieving high accuracy. They show through an extensive comparison, that J48 and random forests were the best at detecting code smells. 
\citet{pecorelli2020large} study the role of different class imbalance mitigating solutions for 11 code smells on 13 software systems. They conclude that balancing classes does not significantly improve performance. However, more recently, \citet{yedida2021value} showed that for defect prediction, using hyper-parameter optimization combined with a novel fuzzy sampling technique significantly improved classification performance. We are motivated by their success to try this approach for code smell detection.

\citet{liu2019deep} use deep learning to detect code smells. Studying the same four code smells as above, they use different deep learning architectures (we will define this term in Section \ref{sec:dl}) for each code smell. For all code smells, they preprocess text features using word2vec \cite{mikolov2013efficient, mikolov2013distributed}. For feature envy, they use convolutional neural networks; for large class detection they use LSTMs; for long method, they use feedforward networks; and for misplaced classes, they use only the word2vec model. We believe it should be possible to detect all code smells using a single model type.

Based on our literature review, we assert that \citet{liu2019deep}  is the  prior state-of-the-art in code smell detection. Further, it is published in a top venue as listed by Google Scholar metrics (TSE), is recent (the latest revision is from September 2021), and is already cited 33 times (the initial publication was in 2019).

\subsection{Feedforward networks}
\label{sec:feedforward}

Feedforward neural networks are a slight extension of multi-layer perceptrons, and are a technology from the 1980s \cite{rumelhart1986learning}. Briefly, a feedforward network is a directed acyclic graph of nodes, where the edges are weighted--these weights form the parameters of the model. At each node, a weighted sum of the inputs is performed, followed by an ``activation function'' to form an output, and weights are updated using the backpropagation algorithm \cite{rumelhart1986learning}. A typical activation function used is the ReLU function, $f(x) = \max(0, x)$, which was introduced by \citet{nair2010rectified}.

Concretely, at a layer $l$, if the weight matrix is denoted as $W^l$, and the constant added (called the ``bias'') is denoted as $b^l$, then the computations at the layer can be summarized as

\[
    a^l = f\left( W^{l}a^{l-1} + b^l \right)
\]

where $a^l$ denotes the ``activations'' at layer $l$.

Despite their simplicity, feedforward networks have been shown theoretically to have large representative capacity; for example, \citet{hornik1989multilayer} shows that these networks can represent any arbitrary decision boundary. This interest is not merely theoretical: \citet{galke2021forget} show that for image classification, feedforward networks can perform competitively with the state-of-the-art algorithms. Nevertheless, modern approaches continue to be more popular in both deep learning and software engineering literature, partly because optimization with modern deep learning layers is easier \cite{santurkar2018does}, and partly because over-parameterized neural networks can still be optimized by simple algorithms such as gradient descent \cite{zou2020gradient, du2018gradient}.

\subsection{Deep learning}
\label{sec:dl}

Deep learning refers to an extension of feedforward networks (or ``artificial neural networks''). Those  had a limited number of hidden layers and used the sigmoid or threshold function at each node (this function is called the ``activation function'' in deep learning literature). They differ from ``deep learners'' in the following respects:

\begin{itemize}
    \item \textbf{Hidden layers:} Modern deep learners typically have many hidden layers, that allow for hierarchical feature selection \cite{zeiler2014visualizing}.
    \item \textbf{Activation functions:} Deep learners now use a variety of activation functions, most notably, the ReLU ($f(x) = \max(0, x)$) \cite{nair2010rectified} function.
    \item \textbf{Architectures:} The ``architecture'' of a deep learner refers to the arrangement of the nodes and the connections between them. In recent times, convolutional neural networks and recurrent neural networks \cite{hochreiter1997long}. Crucially, the ``feedforward'' network architecture refers to the standard multi-layer perceptron setup, but is typically implemented with more layers and the ReLU activation fuinction.
\end{itemize}

Importantly for the context of this paper, however, it is noteworthy that the more recent learners are more complex in that they have orders of magnitude more parameters than prior, simple, feedforward networks (which have been around for decades). For example, a modern deep learner for language modeling can have billions of parameters \cite{radford2019language}. Consequently, these are significantly slower to optimize, although they typically achieve better results.  As we will see in the next section, the more complex networks dominate the SE field as well, and the feedforward network is rather abandoned.

\subsubsection{Autoencoders}

\begin{figure}
    \centering
    \includegraphics[width=\linewidth]{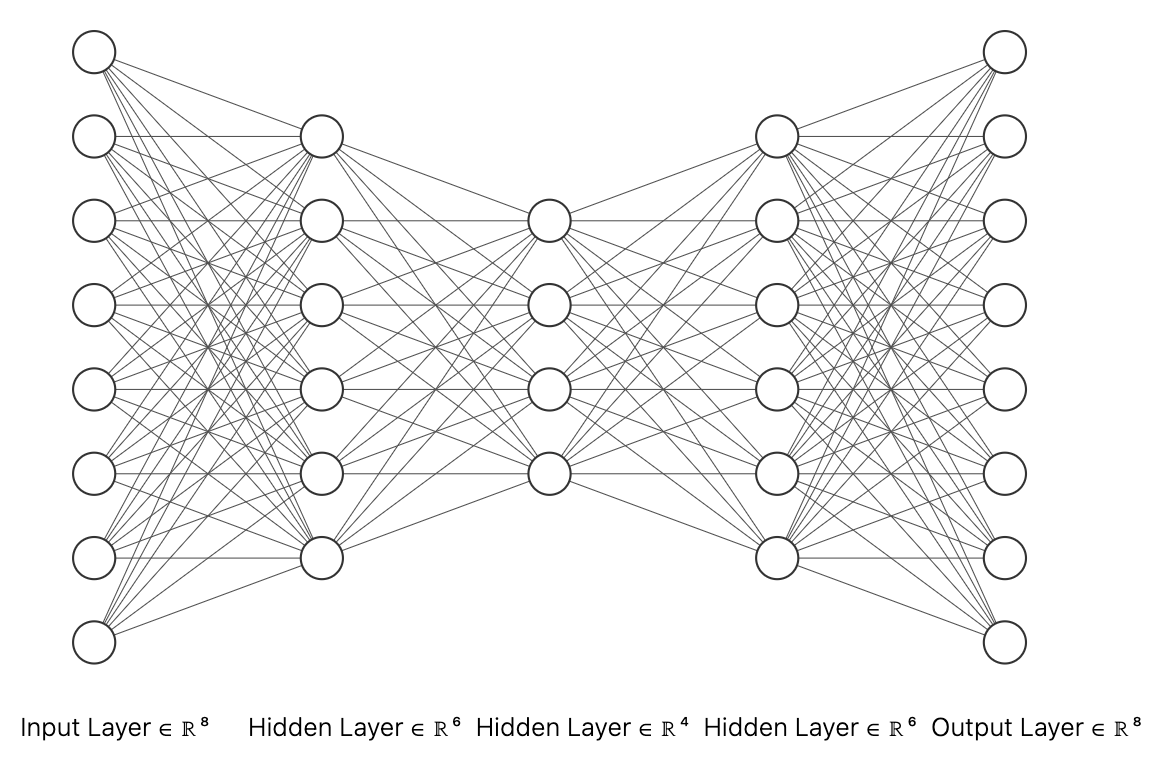}
    \caption{A basic autoencoder. Here, the input of size 8 is reduced to the bottleneck layer of size 4, and the network then increases back to size 8.}
    \label{fig:autoencoder}
\end{figure}

An autoencoder is an encoder-decoder architecture that is used to compress an input into fewer dimensions. While several variations exist, such as the variational autoencoder, in this paper, we only discuss the standard model.

In this model, the network is designed with hidden layers that decrease in size to a certain point (called the \textit{bottleneck} layer), and then increase back to the original input length (see Figure \ref{fig:autoencoder}). The network is then trained to recreate the original inputs, by minimizing the mean squared error (MSE) loss function. This forces the network to learn a mapping from the original input length to the bottleneck length (this part of the network forms the \textit{encoder}), and another mapping from the bottleneck layer to the original input length (called the \textit{decoder}). These can then be used to reduce dimensionality of some data, or recreate the original data from the lower-dimensional embeddings.

\section{Method}
\label{sec:method}

In this section, we discuss our experimental methods in detail.

\subsection{Data}
\label{sec:data}

We use the datasets provided by \citet{liu2019deep}. Briefly, they generate the data for different applications by applying \textit{smell-inducing refactoring}, i.e., a set of refactoring steps that induce code smells in an otherwise well-designed application. For example, moving a method from one class (where it should be) to another is expected to induce \textit{feature envy}.

To generate their training data, positive examples are generated as follows: a set of potential smell-inducing refactorings is constructed, and from this set, sampling without replacement is performed to generate positive examples for each type of code smell. For negative examples, \textit{software entities} (i.e., methods for feature envy and long method, and classes for misplaced class and large class) are considered, and sampling without replacement is done.

\begin{table}[]
    \centering
    \caption{Applications used for this study. Data is taken from \citet{liu2019deep}.}
    \label{tab:datasets}
    \begin{tabular}{lllll}
        \toprule
        Application & Version & \#classes & \# methods & \#samples \\
        \midrule
        JUnit & 4.10 & 123 & 866 & 80,984 \\
        PMD & 5.2.0 & 250 & 2,097 & 80,680 \\
        JExcelAPI & 2.6.12 & 424 & 3,118 & 71,116 \\
        Areca & 7.4.7 & 473 & 5,055 & 76,364 \\
        Freeplane & 1.3.12 & 787 & 6,938 & 74,408 \\
        jEdit & 4.5.0 & 513 & 5,964 & 68,352 \\
        Weka & 3.9.0 & 1,348 & 20,182 & 59,036 \\
        AbdExtractor & 20140630 & 1,695 & 12,608 & 75,156 \\
        Art of Illusion (AoI) & 3.0 & 492 & 6,188 & 67,268 \\
        Grinder & 3.6 & 502 & 3,037 & 78,588 \\
        \bottomrule
    \end{tabular}
\end{table}

The subject applications are listed in Table \ref{tab:datasets}.

\subsection{Statistics}
\label{sec:stats}

Because deep learners are stochastic learners by nature, it is important to compare the distributions of their performance rather than single points. For this reason, to compare results, we use distribution statistics. Specifically, we use the Scott-Knott test using the Cliff's delta effect size test. Briefly, Scott-Knott is a recursive bi-clustering algorithm that seeks to maximize the difference between the means of the resulting groups. Much prior work in SE has used this test for comparison \cite{agrawal2019dodge,agrawal2018better,menzies2018500+}. In particular, if group $l$ is split into $m$ and $n$, Scott-Knott maximizes

\[
    \mathbb{E}[\Delta] = \frac{|m|}{|l|}\left(\mathbb{E}[m] - \mathbb{E}[l]\right)^2 + \frac{|n|}{|l|}\left(\mathbb{E}[n] - \mathbb{E}[l]\right)^2
\]

For Cliff's delta, we use the effect size of 0.147 (small) from \citet{hess2004robust}, so that if Scott-Knott reports that treatment $A$ is better than treatment $B$, then it is different by a non-trivial amount.

\subsection{Reproducing prior work}

The work of \citet{liu2019deep} is open-sourced along with their data; therefore, we were able to reproduce their results. However, due to a different experimental setup used (see Section \ref{sec:setup}), our results are different from the ones claimed in their paper. We report our results in Table \ref{tab:results}.

Notably, the prior work used different sets of architectures for each of the code smells. However, our approach uses the same architecture across the different code smells.

\subsection{GHOST}
\label{sec:ghost}

GHOST (\textbf{G}oal-oriented \textbf{H}yper-parameter \textbf{O}ptimization for \textbf{S}calable \textbf{T}raining) is the fuzzy-oversampling technology proposed by \citet{yedida2021value} for defect prediction. Principally, this relies on feedforward neural networks, but augments them with several modifications that work together to achieve state-of-the-art performance:

\begin{algorithm}[!b]
    \SetAlgoLined
    \For{each sample $\textbf{x}$ in the minority class $c_0$}{
        \For{$i \in \{ 0, 1,\ldots\}$ such that $\frac{(1/n)}{2^i} \geq 1$}{
            Add $(\textbf{x} \pm i\Delta r, c_0)$ to the training set $\lfloor\frac{1/n}{2^i}\rfloor$ times\;
        }
    }
    \caption{Fuzzy sampling}
    \label{alg:wfo}
\end{algorithm}
\begin{algorithm}[!b]
    \SetAlgoLined
    \SetKwInOut{KwInput}{Input}
    \SetKwInOut{KwOutput}{Output}
    \KwInput{dataset $D$,\\ performance threshold $\tau = 0.5$, twoSample = \textbf{false}}
    \KwOutput{optimal hyper-parameters $\theta^*$, performance scores $\boldsymbol \phi$}
    Separate $D$ into train and test sets\;
    \If{twoSample} {
        Apply fuzzy sampling to minority class, reversing the class imbalance\;
    }
    Apply fuzzy sampling to the training set\;
    Apply SMOTE to the resulting training set\;
    Choose a set of key hyperparameters and pre-processors\;
    Use DODGE to perform hyper-parameter optimization over the space of configurations, to obtain optimal hyper-parameter set $\theta^*$ with performance $\phi$\;
    \If{$\boldsymbol \phi < \tau$ \textbf{and} twoSample = \textbf{false}}{
        Run GHOST with twoSample = \textbf{true}\;
    }
    \KwRet{$\theta^*$}
    \caption{GHOST}
    \label{alg:ghost}
\end{algorithm}

\begin{itemize}
   
    \item \textbf{Fuzzy sampling:} \citet{yedida2021value} propose a novel ``fuzzy sampling'' technique that adds points concentrically outwards from each of the minority class points. In doing so, they create a wall of points around each minority sample. They argue that this method of oversampling pushes the decision boundary away from these points, making the classifier more robust to false alarms.
    
    Algorithm \ref{alg:wfo} shows the fuzzy sampling algorithm from the original paper. In that algorithm, $n$ is the fraction of samples belonging to the minority class (i.e., the class imbalance ratio), $c_0$ is the minority class, $\Delta r$ is a user-specified parameter. We use the default suggested in the original paper of 0.01.
    
    In their paper, they show that while doing this once is effective, doing it twice can yield even better results. This is because fuzzy sampling inverts the class imbalance, so that the minority class becomes the majority. The authors perform fuzzy sampling a second time to make the classifier robust to false \textit{negatives}, and finally balance out the imbalance with SMOTE \cite{Chawla02}, which adds synthetic points halfway between minority samples and their near neighbors. This technique, which they call \textit{twoSample} in their paper, yields results with excellent recall and precision, and low false alarm rates.
     \item \textbf{Hyper-parameter Optimization:} Many learning algorithms and preprocessors come with multiple parameters that are set with engineering judgement; such parameters are called hyper-parameters. However, it is difficult to judge what values of these parameters will work well in practice; therefore, it is important to find optimal values of these hyper-parameters: this search is called hyper-parameter optimization.
     
     GHOST relies on the DODGE \cite{agrawal2019dodge} hyper-parameter optimizer. In this paper, we also use the DODGE hyper-parameter optimizer to make this work a direct extension of their paper. As noted in their paper, a recent study \cite{agrawal2021simpler} showed that DODGE is better for SE datasets, which have low intrinsic dimensionality.
    
    DODGE is a tabu search-based hyper-parameter optimization algorithm. Specifically, it uses the heuristic that if two configurations yield performance scores within some $\epsilon$ of each other, then configurations close to them should not be explored further (called the ``$\epsilon$-domination rule''). In this way, DODGE partitions the configurations based on the metric space, using this $\epsilon$-domination rule.
    
\end{itemize}

The overall GHOST system is summarized in Algorithm \ref{alg:ghost}, which is a shortened version of the algorithm from the original paper. The algorithm starts by applying fuzzy sampling to the training set (Line 5)--it is here that an additional fuzzy sampling step may be applied (lines 2-4)--followed by SMOTE (line 6). Having pre-processed the data in this way, the user-defined set of hyper-parameter configurations is collected (line 7), and passed to DODGE for tuning. DODGE is set to run for 30 iterations as recommended by its authors \cite{agrawal2019dodge}, and it returns an optimal set of hyper-parameters, with which a performance $\phi$ is obtained (line 8). If this performance is below some threshold $\tau$, we re-run it with \textit{twoSample} set to true (lines 9-11): this allows for efficient results, since this option significantly increases the size of the training set, therefore increasing the runtime of the algorithm.

\subsection{Experimental Setup}
\label{sec:setup}

While a successful prototype  in its home domain, GHOST's ``fuzzy sampling'' was only ever tested on defect prediction. Hence, here, we check the generality of that method with a new case study applied to   code smell detection.

To that end,
we ran all our code on a machine with an RTX 2080 Super and an AMD Epyc Rome CPU. While the code of \citet{liu2019deep} performs cross-project code smell detection (i.e., train on 9 projects and test on the 10th), we use within-project code smell detection (i.e., 20 times, split into train/test of 70-30\%). We do this for the following reasons:

\begin{itemize}
    \item In this study, we are interested in the in-distribution learning power of GHOST, rather than its cross-project generalization ability.
    \item In general, when detecting code smells in an application in say, one file, code from other parts of the application (potentially, with labels) is available. 
    \item In case much other data is not available, pre-training can be leveraged, which has been recently shown in the SE literature to be effective for small datasets \cite{prenner2021making}. However, these datasets are from larger applications with sufficient data to train, and so pre-training was not necessary.
\end{itemize}

Therefore, the results we obtain for their approach are different than the ones in their paper. For this reason, we report in Table \ref{tab:results}, (a) the results from their paper (first set of columns), (b) the results from our reproduction of their work (second set of columns), and (c) our results (third set of columns). Because we do not have 20 repeats of the first set of columns, all Scott-Knott comparisons are between the second and third sets of columns.

For a statistical comparison, we run each experiment 20 times and run a Scott-Knott test as described in  \ref{sec:stats}.

\section{Results}
\label{sec:results}

\begin{table*}
    \centering
    { \small
    \caption{RQ2 results: 
    performance seen with  different code smells. \fcolorbox{black}{lightgray}{Gray cells} indicate better results (comparison is between the second and third sets of columns), using a Scott-Knott test over 20 repeats. In this table ``SOTA'' = the prior state of the art (and comes from~\citet{liu2019deep}).}
    \label{tab:results}
    \begin{subtable}{\linewidth}
        \centering
        \caption{Feature envy }
        \begin{tabular}{l|llll|llll|llll}
            \toprule
            \textbf{Dataset} & \multicolumn{4}{c}{\textbf{SOTA (original)}} & \multicolumn{4}{c}{\textbf{SOTA (reproduced)}} & \multicolumn{4}{c}{\textbf{Fuzzy sampling (with GHOST)}}  \\
             \midrule
             & \textbf{Precision} & \textbf{Recall} & \textbf{F1} & \textbf{AUC} & \textbf{Precision} & \textbf{Recall} & \textbf{F1} & \textbf{AUC} & \textbf{Precision} & \textbf{Recall} & \textbf{F1} & \textbf{AUC} \\ 
             \midrule
            abd-extractor & 29.7 & 76.7 & 42.8 & 74.8 & 79.5 & 80 & 79.7 & 88 & \gray 98 & \gray 98 & \gray 98 & \gray 99.3 \\ 
            aoi30 & 36.7 & 97 & 53.2 & 93.6 & 79.2 & 79.2 & 79.2 & 87.3 & \gray 97.7 & \gray 97.7 & \gray 97.7 & \gray 99.1 \\ 
            areca & 50 & 88 & 63.8 & 91.9 & 80.1 & 79.9 & 80 & 88.3 & \gray 98 & \gray 98 & \gray 98 & \gray 99.3 \\ 
            freeplane & 36.2 & 94.1 & 52.3 & 83.1 & 80.6 & 80.4 & 80.5 & 88.8 & \gray 98.2 & \gray 98.2 & \gray 98.2 & \gray 99.3 \\ 
            grinder & 31.2 & 88.6 & 46.1 & 85.2 & 80 & 80.1 & 80 & 88.1 & \gray 97.7 & \gray 97.7 & \gray 97.7 & \gray 99.1 \\ 
            jedit & 38.2 & 91.3 & 53.9 & 84.3 & 78.7 & 78.5 & 78.6 & 87 & \gray 97.8 & \gray 97.8 & \gray 97.8 & \gray 99.2 \\ 
            jexcelapi & 34 & 88.9 & 49.2 & 90 & 78.5 & 78.2 & 78.3 & 86.3 & \gray 97.7 & \gray 97.7 & \gray 97.7 & \gray 99.2 \\ 
            junit & 50 & 82.2 & 62.1 & 85.7 & 80.3 & 80.2 & 80.2 & 88 & \gray 97.7 & \gray 97.7 & \gray 97.7 & \gray 99.1 \\ 
            pmd & 37.4 & 86 & 52.1 & 84.9 & 79.8 & 79.6 & 79.7 & 87.8 & \gray 97.8 & \gray 97.8 & \gray 97.8 & \gray 99.2 \\ 
            weka & 38.2 & 87 & 53.1 & 78.7 & 77.9 & 77.5 & 77.7 & 85.8 & \gray 97.8 & \gray 97.8 & \gray 97.8 & \gray 99.1 \\ 
            \bottomrule
        \end{tabular}
    \end{subtable}
    \begin{subtable}{\linewidth}
        \centering
        \caption{Long Method }
        \begin{tabular}{l|llll|llll|llll}
            \toprule
            \textbf{Dataset} & \multicolumn{4}{c}{\textbf{SOTA (original)}} & \multicolumn{4}{c}{\textbf{SOTA (reproduced)}} & \multicolumn{4}{c}{\textbf{Fuzzy sampling (with GHOST)}}  \\
             \midrule
             & \textbf{Precision} & \textbf{Recall} & \textbf{F1} & \textbf{AUC} & \textbf{Precision} & \textbf{Recall} & \textbf{F1} & \textbf{AUC} & \textbf{Precision} & \textbf{Recall} & \textbf{F1} & \textbf{AUC} \\ 
             \midrule
             areca & 42.7 & 73.8 & 54.1 & 78.5 & 41.7 & \gray 77.8 & \gray 52.5 & \gray 72.2 & 41.7 & 70 & 51.5 & 71.2 \\ 
            freeplane & 46.4 & 75.6 & 57.5 & 78.8 & 44.6 & 78.2 & 55.6 & 70.3 & \gray 45.8 & 76.7 & \gray 57 & \gray 71.1 \\ 
            jedit & 52.2 & 83.5 & 64.2 & 77.2 & 53.4 & 82.5 & \gray 63.4 & \gray 71.3 & \gray 54.1 & 80.5 & 62.2 & 70.4 \\ 
            junit & 58.5 & 52.9 & 55.6 & 72.6 & 51.7 & 67.1 & 56 & 66.1 & 51.3 & 64.9 & 54.1 & 61.6 \\ 
            pmd & 37.1 & 70.6 & 48.6 & 77.4 & 35.2 & 84.8 & 47.6 & 73.6 & 35.4 & 77.8 & 47.1 & 72.9 \\ 
            weka & 50.2 & 79.3 & 61.5 & 81.8 & \gray 49.9 & 83.9 & \gray 62.3 & \gray 76.7 & 47.7 & \gray 86.3 & 61.2 & 76.1 \\ 
            abd-extractor & 32.3 & 80.6 & 46.2 & 78.8 & 39.1 & 73.5 & 50.7 & 73.4 & 38.9 & 73.6 & 50.5 & 73.2 \\ 
            grinder & 37.2 & 71.8 & 49 & 74.1 & 36.1 & 66.3 & 47.5 & 65.9 & 38 & 73.9 & 46.2 & 64.8 \\ 
            aoi30 & 37.6 & 87.6 & 52.6 & 80.3 & 38.4 & 81.2 & 51.8 & 72.6 & 38.2 & 81.5 & 52.2 & 72.7 \\ 
            jexcelapi & 32.6 & 83.6 & 46.9 & 88.5 & 38.5 & \gray 81.5 & \gray 52.1 & \gray 80.3 & 37.3 & 81.1 & 50.9 & 79.8 \\ 
            \bottomrule
        \end{tabular}
    \end{subtable}
    \begin{subtable}{\linewidth}
        \centering
        \caption{Large Class }
        \begin{tabular}{l|llll|llll|llll}
            \toprule
            \textbf{Dataset} & \multicolumn{4}{c}{\textbf{SOTA (original)}} & \multicolumn{4}{c}{\textbf{SOTA (reproduced)}} & \multicolumn{4}{c}{\textbf{Fuzzy sampling (with GHOST)}}  \\
             \midrule
             & \textbf{Precision} & \textbf{Recall} & \textbf{F1} & \textbf{AUC} & \textbf{Precision} & \textbf{Recall} & \textbf{F1} & \textbf{AUC} & \textbf{Precision} & \textbf{Recall} & \textbf{F1} & \textbf{AUC} \\ 
             \midrule
             areca & 11.4 & 80 & 20 & 68.8 & 33.7 & 78.3 & 47.1 & 50 & \gray 82.9 & 77.5 & \gray 80.1 & \gray 89.5 \\ 
            freeplane & 12 & 70 & 20.4 & 72.8 & 6.1 & \gray 100 & \gray 11.5 & 50 & 0 & 0 & 0 & \gray 68.8 \\ 
            jedit & 15 & 75 & 25 & 77.4 & 61.5 & \gray 100 & 76.2 & 50 & \gray 76.8 & 81.8 & 79.2 & \gray 75 \\ 
            junit & 11.8 & 40 & 18.2 & 71.8 & 11.9 & 100 & 21.3 & 50 & \gray 32.2 & 71.4 & \gray 44.4 & 74.6 \\ 
            pmd & 16.7 & 100 & 28.6 & 83.5 & 11.4 & \gray 100 & 20.5 & 50 & \gray 23.8 & 23.6 & 23.7 & \gray 63.1 \\ 
            weka & 10 & 94.4 & 18.2 & 68.6 & 0 & 0 & 0 & 50 & \gray 42.9 & \gray 71.4 & \gray 53.6 & \gray 70.7 \\ 
            abd-extractor & 16.5 & 79.4 & 27.3 & 79.8 & 9.1 & 14.3 & 11.1 & 50 & \gray 27.9 & 28.6 & \gray 28.2 & \gray 65.3 \\ 
            grinder & 12.7 & 70 & 21.5 & 79.2 & 0 & 0 & 0 & 50 & 0 & 0 & 0 & 50 \\ 
            aoi30 & 12.3 & 81.8 & 21.4 & 78.5 & 30 & 100 & 46.2 & 50 & \gray 53.8 & 80 & \gray 64.3 & \gray 75.7 \\ 
            jexcelapi & 22 & 84.6 & 34.9 & 80.9 & 15.7 & \gray 100 & 27.1 & 50 & 57.7 & 73.5 & 64.6 & 80.4 \\ 
            \bottomrule
        \end{tabular}
    \end{subtable}
    \begin{subtable}{\linewidth}
        \centering
        \caption{Misplaced Class }
        \begin{tabular}{l|llll|llll|llll}
            \toprule
            \textbf{Dataset} & \multicolumn{4}{c}{\textbf{SOTA (claimed)}} & \multicolumn{4}{c}{\textbf{SOTA (reproduced)}} & \multicolumn{4}{c}{\textbf{GHOST}}  \\
             \midrule
             & \textbf{Precision} & \textbf{Recall} & \textbf{F1} & \textbf{AUC} & \textbf{Precision} & \textbf{Recall} & \textbf{F1} & \textbf{AUC} & \textbf{Precision} & \textbf{Recall} & \textbf{F1} & \textbf{AUC} \\ 
             \midrule
             areca & 93.6 & 92.6 & 93.1 & 99.8 & 34.1 & 100 & 50.9 & 98.6 & \gray 44.7 & 100 & \gray 61.8 & \gray 100 \\ 
            freeplane & 94.9 & 91.5 & 93.2 & 99.8 & 32.7 & 100 & 49.3 & 96.3 & \gray 43.3 & 99.4 & \gray 60.3 & \gray 99.4 \\ 
            jedit & 55 & 100 & 71 & 99.2 & 32.6 & 100 & 49.2 & 98.7 & \gray 36 & 100 & \gray 53 & \gray 100 \\ 
            junit & 41.4 & 100 & 58.5 & 99.5 & 35.2 & 96.1 & 51.6 & 93.4 & \gray 36.3 & \gray 96.3 & 52.7 & \gray 97.3 \\ 
            pmd & 56.1 & 97.9 & 71.3 & 99.6 & 41.2 & 96.8 & 57.7 & 92 & \gray 53.8 & \gray 99.3 & \gray 69.8 & \gray 99.6 \\ 
            weka & 96.8 & 94.2 & 95.5 & 99.8 & 48.5 & 99.8 & 65.3 & 99.7 & 48.4 & \gray 99.9 & 65.2 & \gray 99.8 \\ 
            abd-extractor & 95.1 & 92.3 & 93.7 & 99.7 & \gray 44.5 & 99.7 & \gray 61.6 & 99.5 & 43.8 & \gray 100 & 60.9 & \gray 99.8 \\ 
            grinder & 92.5 & 82.2 & 87.1 & 98.4 & 18.2 & 90.6 & 30.4 & 85.6 & \gray 24.5 & \gray 100 & \gray 39.4 & \gray 100 \\ 
            aoi30 & 76.4 & 100 & 86.6 & 99.4 & 43.6 & 100 & 60.7 & 92.2 & 44 & 100 & 61.8 & \gray 100 \\ 
            jexcelapi & 39.6 & 90.5 & 55.1 & 92.5 & 21.6 & 86.4 & 34.6 & 74.3 & \gray 32.6 & \gray 96.1 & \gray 48.6 & \gray 96.3 \\ 
            \bottomrule
        \end{tabular}
    \end{subtable}
    }
\end{table*}

\begin{table*}
    \centering
    \caption{Summary of Table \ref{tab:results} for each code smell.}
    \label{tab:summary}
    \begin{tabular}{l|llll|l}
        \toprule
         & \textbf{Precision} & \textbf{Recall} & \textbf{F1} & \textbf{AUC} & \textbf{Total} \\
        \midrule
        \multicolumn{6}{c}{Feature Envy} \\
        \midrule
        win & \sbar{100} {10} & \sbar{100} {10} & \sbar{100} {10} & \sbar{100} {10} & \sbar{100} {40} \\
        tie & \sbar{0} {0} & \sbar{0} {0} & \sbar{0} {0} & \sbar{0} {0} & \sbar{0} {0} \\
        loss & \sbar{0} {0} & \sbar{0} {0} & \sbar{0} {0} & \sbar{0} {0} & \sbar{0} {0} \\
        \midrule
        \multicolumn{6}{c}{Long Method} \\
        \midrule
        win & \sbar{20} {2} & \sbar{10} {1} & \sbar{10} {1} & \sbar{10} {1} & \sbar{13} {5} \\
        tie & \sbar{70} {7} & \sbar{70} {7} & \sbar{50} {5} & \sbar{50} {5} & \sbar{60} {24} \\
        loss & \sbar{10} {1} & \sbar{20} {2} & \sbar{40} {4} & \sbar{40} {4} & \sbar{27} {11} \\ 
        \midrule
        \multicolumn{6}{c}{Large Class} \\
        \midrule
        win & \sbar{80} {8} & \sbar{10} {1} & \sbar{60} {6} & \sbar{70} {7} & \sbar{55} {22} \\
        tie & \sbar{20} {2} & \sbar{50} {5} & \sbar{30} {3} & \sbar{30} {3} & \sbar{33} {13} \\
        loss & \sbar{0} {0} & \sbar{40} {4} & \sbar{10} {1} & \sbar{0} {0} & \sbar{12} {5} \\
        \midrule
        \multicolumn{6}{c}{Misplaced Class} \\
        \midrule 
        win & \sbar{70} {7} & \sbar{60} {6} & \sbar{60} {6} & \sbar{100} {10} & \sbar{73} {29} \\
        tie & \sbar{20} {2} & \sbar{40} {4} & \sbar{30} {3} & \sbar{0} {0} & \sbar{23} {9} \\
        loss & \sbar{10} {1} & \sbar{0} {0} & \sbar{10} {1} & \sbar{0} {0} & \sbar{20} {2} \\
        \bottomrule 
    \end{tabular}
\end{table*}

We discuss our results in the context of each research question. Before that, we discuss the results from our literature review.

\subsection{Generalizing ability of GHOST}

In this section, we discuss the answer to \textbf{RQ1}, which was, ``\textit{Can GHOST achieve state-of-the-art results in code smell detection?}''. A more thorough study across more domains is left as future work.

\begin{table}[h]
    \centering
    \caption{Percentage of values in the distance matrix (for feature envy) that were 0. Note that the percentages are extremely small (e.g. for abd-extractor, $0.0003\%$ of values were 0).}
    \label{tab:zeros}
    \begin{tabular}{ll}
        \toprule
        \textbf{Dataset} & \textbf{\% of 0s} $\times 10^{5}$ \\
        \midrule
        abd-extractor & 33.2 \\
        aoi30 & 47.3 \\
        areca & 36 \\
        freeplane & 38.2 \\
        grinder & 33.4 \\
        jedit & 16.3 \\
        jexcelapi & 40.9 \\
        junit & 32.8 \\
        pmd & 31.9 \\
        weka & 58 \\
         \bottomrule
    \end{tabular}
\end{table}

The results of GHOST on code smell detection are shown in Table \ref{tab:results}, which are summarized in Table \ref{tab:summary}. For feature envy, GHOST is better \textit{all} the time, reaching near-perfect scores. Because the scores were so high, we took the extra step to ensure that there was no data leakage (i.e., none of the samples from the training set leaked into the test set). To check this, we computed the distance matrix (using the Euclidean distance) between the train and test sets, and checked for values of 0 (which would happen if the points were the same). The total number of values in the distance matrix ranged from \textasciitilde 800 million to \textasciitilde 1.3 billion (after oversampling, which increases the size of the training set), and of these, between 2,000 to 4,000 were 0. This means that of $\sim\sqrt{10^9} = \sim 30,000$ samples, $\sim \sqrt{10^3} = \sim 30$ were the same. The exact percentages of values that were 0 are shown in Table \ref{tab:zeros}. Note that these values are \textit{extremely small}, e.g. for abd-extractor, $33.2 \times 10^{-5}\% = 0.0003\%$ of values were zero. We argue that this had no effect on the performance of the learner on the test set.

On long method detection, we generally lose (11 times), although most of the time, we tie (24 times). We argue that this is because \citet{liu2019deep} use feedforward networks (their code uses the \texttt{MLPClassifier} class from \texttt{sklearn}) for their long method detection as well, and therefore there is not much performance gain to be expected.

Our wins continue, however, on large class and misplaced class detection. On the former, we win $22/40=55\%$ of the time, and tie $13/40=32.5\%$ of the time. Our wins are greater still in the latter case (misplaced class), where we win $29/40=72.5\%$ of the time and tie $9/40=22.5\%$ of the time. 

The above results strongly favor GHOST over prior work. Note that these results are from 20 repeats, with a statistical test to determine ``better''. Therefore, we have sufficient evidence to say:

\begin{formal}
    \noindent
    GHOST also achieves state-of-the-art performance on code smell detection.
\end{formal}

\subsection{Why do feedforward networks work so well?}

In this section, we discuss the answer to \textbf{RQ2}, which was, ``\textit{Why do feedforward networks work so well?}''

There are several theories on the generalizing capabilities of feedforward networks in general:

\begin{enumerate}[(a)]
    \item \citet{hornik1989multilayer} show, using prior work \cite{cybenko1989approximation}, that feedforward neural networks with as few as one hidden layer, with a sufficient number of units, are universal approximators.
    \item More recently, \citet{jacot2018neural} showed that a feedforward learner with infinite width (the number of units in the layer)\footnote{In practice, this translates to ``sufficient width''.} is a linear model under a kernel they called the ``neural tangent kernel'', and that it can approximate any arbitrary function.
    \item \citet{montufar2014number} showed that the decision boundary of neural networks are piecewise linear, and that there is a derivable lower and upper bound on the number of linear pieces constituting the boundary for a given network. This idea was exploited by \citet{yedida2021value} to design the structure of their networks.
    \item \citet{galke2021forget} show that several tasks for which complex deep learners have been applied recently can also be done with feedforward networks with no loss of performance.
    \item \citet{yedida2021value} showed that using novel preprocessing methods, one can push the decision boundary away from points, making the classifier more robust to noise.
\end{enumerate}

From the above works, we derive the following important lessons:

\begin{enumerate}[(a)]
    \item The deep learning literature broadly agrees that feedforward networks can approximate any arbitrary function \cite{cybenko1989approximation, hornik1989multilayer, montufar2014number}.
    \item The decision boundary is known to be piecewise-linear, but also ``malleable'' (i.e., its shape can be changed using appropriate preprocessing). It has been shown \cite{hornik1989multilayer} that the composition of the nonlinearities in feedforward networks allows for arbitrary decision boundary shapes.
\end{enumerate}
Based on these examples we were motivated to see if  there is a  general pattern or test that lets us recommend using (or avoiding) GHOST.
While any such test can only be a heuristic (since it is difficult to predict the performance of a deep learner in general), we offer one such heuristic that seems suitable for future research.

We start by designing a simple autoencoder as follows:
\bi
\item
Let $2^k$ be the highest power of 2 that is lesser than the input vector length.
\item
Fix the bottleneck layer with engineering judgement (for SE tasks, which are known to be simpler than general AI tasks \cite{agrawal2021simpler}, we found 32 or 64 to be useful; for general AI tasks, we use 128\footnote{Using 64 did not change the results we show below.}).
\item
Then, in the encoder, set the number of units to be $k, k/2, k/4, \ldots$ until the bottleneck layer.
\item
Design the decoder as the mirrored version of the encoder.
\ei
(For clarity, we will give an example of this towards the end of this section.)

As is usual with the standard autoencoder, train this neural network on the input data with the mean squared error (MSE) loss. Our heuristic is then:

\begin{formal}
    If the MSE loss of the autoencoder designed as discussed above is below 1,000, attempt a feedforward network before trying more complex methods.
\end{formal}

It is worth noting that autoencoders, like any deep learner, can get stuck in local optima because of the non-convex nature of loss functions \cite{choromanska2015loss}. It is worth attempting this three times to ensure that a loss higher than 1,000 is because it is not possible, rather than poor optimization.

We tested this heuristic on the original defect prediction datasets used by \citet{yedida2021value}, and it was true for \textit{all} of them. Furthermore, this heuristic was also true for the code smell detection datasets of this paper. \BLUE However, it does not suffice to say the heuristic worked for our datasets; we need to show cases where (a) it is not true, and (b) the simpler network failed to make a valid heuristic \BLACK. To do this, we pulled 3 common image classification datasets: MNIST (digit classification), CIFAR-10 (10-class image classification), and CIFAR-100 (100-class image classification).

For example, consider the MNIST dataset. This consists of black-and-white images of digits (0-9), sized 28 x 28. We flatten these images to one vector of length 784. Then, the highest power of 2 lower than this is 512, and we choose the bottleneck layer to be 128 units; therefore, our autoencoder architecture is 784 (input) - 512 - 256 - 128 (bottleneck) - 256 - 512 - 784. 

To our surprise, for the MNIST dataset, the heuristic was true; on applying a feedforward network (without hyper-parameter optimization) to it, we achieved a classification accuracy of 96.3\%--not state-of-the-art, but respectable, and would be improved with the hyper-parameter optimization of GHOST.

For both the CIFAR-10 and CIFAR-100 datasets, the heuristic was false; indeed, the feedforward network fell significantly short of convolutional neural networks that are typically applied to these datasets: on CIFAR-10, we achieved 44\% accuracy (modern networks can achieve 90+\% accuracy), and on CIFAR-100, we achieved merely 18.3\% (while a modern network would achieve 75+\% accuracy). Therefore, we have (a) case studies where the heuristic is false and our simpler networks failed (as expected by the result of the heuristic), and also (b) case studies where the heuristic is true, and the feedforward network was sufficient. Our experiments on the defect prediction datasets, code smell detection datasets, and image classification datasets show that this heuristic is reliable.

These experiments seem to reinforce the claim made by \citet{agrawal2021simpler} who said that software analytics may be simpler than standard AI datasets. If a standard autoencoder cannot compress the input data into the 64 dimensions of its bottleneck layer, it may mean that those image datasets (where it failed) are intrinsically much higher dimensional. However, our heuristic's success on both the code smell detection and defect prediction data suggest that it should be applied as a test for whether feedforward networks should be used instead of more complex approaches.

\section{Broader Implications}
\label{sec:broader}

This paper is not the first to doubt the ubiquitous benefits of DL for SE. In \citet{yedida2021value} it was shown that with some preprocessing and hyper-parameter optimization, feedforward networks can achieve state-of-the-art performance for defect prediction. 

But when should we use deep learning? We posit here that when dealing with very high-dimensional data, such as images, it is worth using deep learning (see the previous section); indeed, this has been done previously in SE (see \citet{chen2018ui}). Another case when deep learning would be useful in SE is for extracting meaningful embedding vectors from code; see \citet{alon2018code2seq} and \citet{peng2021integrating} for recent approaches. Our heuristic above was able to distinguish when feedforward networks might succeed on three domains (image classification, defect prediction, code smell detection), and we believe that this distinguishing power would be useful to the community.

While the obvious advancement provided by this work is in a new state-of-the-art system for code smell detection, there is another underlying contribution. Specifically, this work adds to a small but growing interest in the revitalization of feedforward networks and their potential. For example, as discussed in the previous section, \citet{jacot2018neural}, in 2018, visited the idea of infinite-width but shallow feedforward networks, and \citet{galke2021forget} in 2021 showed that a lot of tasks for which more advanced models have been applied can also be solved, with little loss of performance, using feedforward networks.

\begin{figure}[!t]
    \centering
    \includegraphics[width=.8\linewidth]{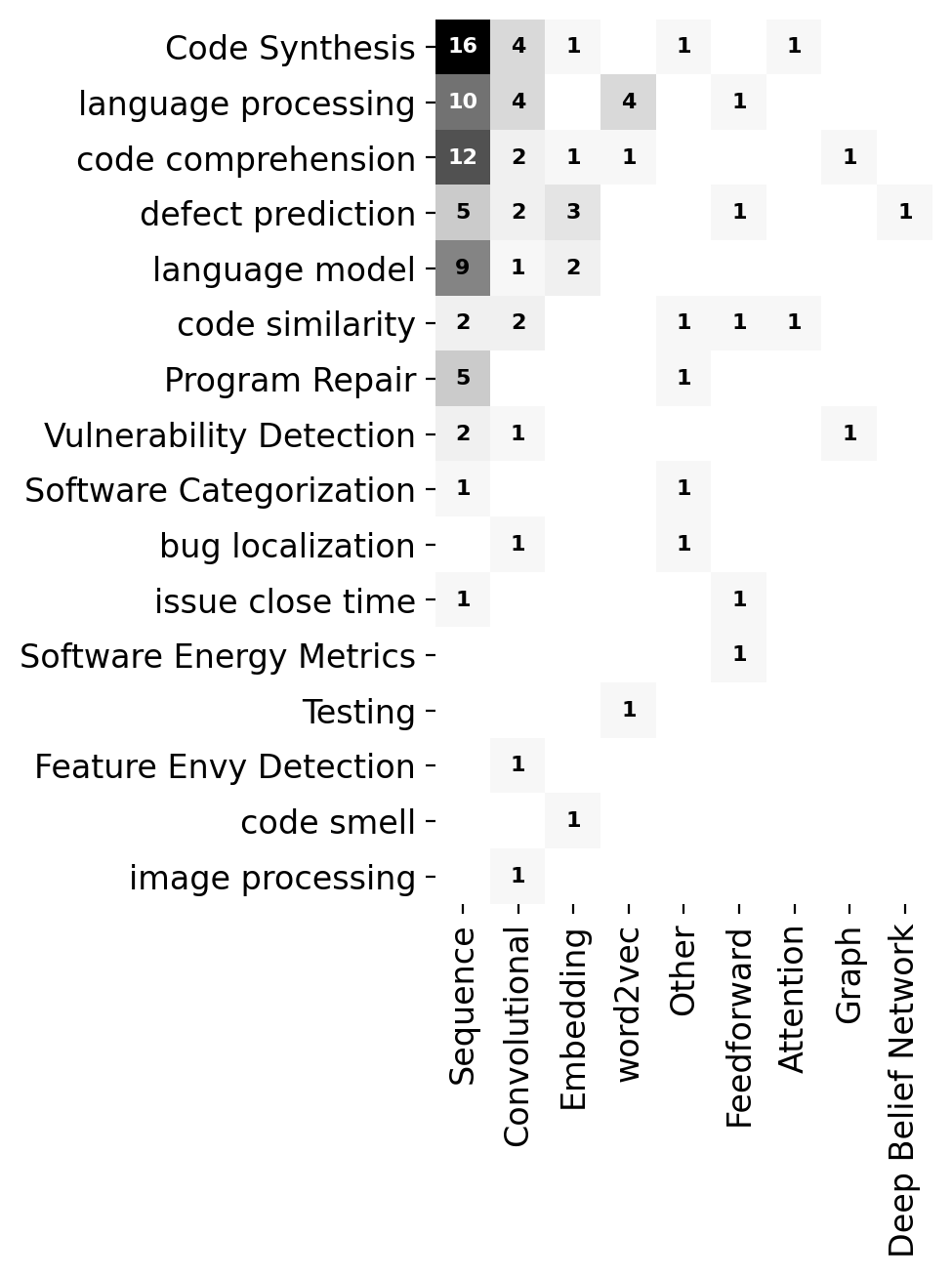}
    \caption{State of deep learning in software engineering. For details on how this chart
    was generated, see \S\ref{sec:discussion}}
    \label{fig:dl4se}
\end{figure}

While this body of literature may have some renewed interest, overall, Figure \ref{fig:dl4se} shows that, at least in software engineering, this interest is minimal. Further, that diagram does not illustrate that often, when feedforward learners \textit{are} used, they are used as part of a larger deep learning system that include convolutional or recurrent networks to extract features, and a few feedforward layers at the end to make predictions. However, we argue for \textit{pure} feedforward networks, i.e., no convolutional or recurrent models. This leads to simpler models that run faster. We now discuss yet another motivation for using feedforward networks: knowledge distillation.


\textit{Knowledge distillation} \cite{hinton2015distilling} is a technique that allows a ``teacher'' model to train a ``student'' model. \BLUE Specifically, a large number of inputs is generated and fed to the originally trained, more complex model (a deep learner), and the outputs are captured. These input-output pairs are used to train a simpler, often faster or more interpretable model (such as a smaller deep learner or a decision tree). \BLACK Surprisingly, the student model often matches or even outperforms the teacher model. Knowledge distillation has been widely studied \cite{phuong2019towards,park2019relational,kim2016sequence}, and shown to be useful for various tasks \cite{chen2017learning,gou2021knowledge}. Knowledge distillation also has the benefits of yielding simpler, faster, models, and being amenable to transfer learning \cite{yim2017gift}.

However, there have been shown to be limitations to knowledge distillation \cite{Cho_2019_ICCV}. Specifically, a key insight of \citet{Cho_2019_ICCV} is that larger models may not be effective teachers, especially for smaller students, and typical solutions such as multiple rounds of distillation fail. From this, we infer that because convolutional models are typically significantly larger than their feedforward counterparts (due to the additional parameters introduced by the convolutional layers), they may be less effective teachers for decision tree learners. Therefore, we advocate instead for obtaining performance by leveraging hyper-parameter optimization on feedforward networks, which yield fast, simple models; then, if an interpretable model is required, knowledge distillation can be used to obtain a decision tree with similar performance. This is because our feedforward networks are significantly smaller than deep learners such as convolutional neural networks, and so can be effective teacher models in knowledge distillation.

\section{Related Work}
\label{sec:discussion}
In our introduction, this paper was motivated by a concern that deep learning
was being applied in SE without much consideration on how best to apply that technology.  This section expands on that point.

Figure \ref{fig:dl4se} shows the state of deep learning applied in software engineering. We achieved that diagram by performing a systematic literature review as described below:

\begin{itemize}
    \item \textbf{Seed:} Our approach started with collecting relevant papers. As a seed, we collected papers from the recent literature review conducted by Watson ~\cite{watson2020deep}.
    \item \textbf{Search:} To this list, we added papers added by our own searches on Google Scholar. Our search keywords included ``deep learning AND software'', ``deep learning AND defect prediction'', and ``deep learning AND bug''.
    \item \textbf{Filter:} Next, we filtered papers using the following criteria: 
    
        \begin{enumerate}[(a)]
            \item One of the following criteria is met:
                \begin{enumerate}[(i)]
                    \item published in top venues as listed in Google Scholar metrics for Software Systems, Artificial Intelligence, and Computational Linguistics
                    \item released on arXiv in the last 4 years
                    \item widely cited ($>$ 100 cites)
                \end{enumerate}
            \item has at least 10 cites per year, unless it was published in or after 2017 (the last four years)
        \end{enumerate}
    \item \textbf{Backward Snowballing:} As recommended by \citet{wohlin2014guidelines},  we performed ``snowballing'' on our paper (i.e. we added papers cited by the papers in our list that also satisfy the criteria above).
    Our snowballing stopped when  either (a) the list of papers cited by the current generation is a subset of the papers already in the list, or (b) there were no further papers found.
\end{itemize}

This produced a list of 118 papers. Next, we categorized each paper by manually labeling (a) the architecture used, and (b) the task solved. When multiple architectures were used, we listed all. For the task solved, because of the variety, we grouped some tasks together. Specifically, we made the following groups: (a) \textbf{code similarity} refers to both code similarity and code clone detection; (b) \textbf{language processing} refers to any natural language processing or programming language processing task, such as API translation; (c) \textbf{code comprehension} refers to code summarization and code comprehension tasks such as analogous API mining.


This figure reveals a significant skew towards the use of sequence and convolutional (i.e., more complex) models, as opposed to feedforward networks. For example, out of the numbers in Figure \ref{fig:dl4se}, 59\% are Sequence models, and 17\% are Convolutional. That is, 76\% (a little over three-quarters) of the grid is taken up by modern, more complex deep learners. Meanwhile, feedforward networks only occupy 4.5\% of the grid. In this paper, we showed another case study where more complex networks were outperformed by feedforward networks with fuzzy sampling and hyper-parameter optimization. 

We suggest that this paper highlights an issue that is more general than just fuzzy sampling or code smell detection. Rather, we wonder if software analytics is exploring very complex methods without comparing those approaches to simpler alternatives. We are not the first to pose this question: \citet{galke2021forget} achieve competitive results on image classification tasks using feedforward networks, and \citet{menzies2018500+} show that simple tuning of Latent Dirichlet Allocation (LDA) was sufficient to outperform deep learning models.

In an enterprise setting, where developers must meet deadlines for user stories, one cannot expect them to wait for deep learning inference, which may take minutes. Further, it is impractical to assume that all developer machines will have access to GPUs to speed up deep learning computation. Therefore, the more we can speed up such inference, the more useful these tools will be for developers.

\section{Conclusion}
\label{sec:conclusion}

If a method appears to be sweeping a field
(e.g. deep learning in software analytics), it is prudent to check the value
of that method. 
This paper presented a new case study applying the GHOST fuzzy sampling algorithm in code smell detection. 

In summary, while
originally proposed for defect prediction, GHOST outperformed the state-of-the-art in code smell detection. Hence we can say   that the algorithm is general to at least two tasks
(defect prediction and code smell detection). This work also provided a test for when simpler feedforward networks might suffice for a problem, through the use of autoencoders, with experiments on three domains.
We leave it as future work to extend this work further (such as more SE analytics tasks). 

\BLUE
We note here that a lot of this work can be automated; indeed, we have published a package\footnote{\url{https://pypi.org/project/raise-utils/}} containing the GHOST algorithm so that it may be implemented in a few lines of code by practitioners. That said, some feature engineering may be required; this can be static code features, or automated features from systems such as code2vec \cite{alon2019code2vec}. \BLACK

To conclude, 
we find we can comment on the
 following points made in the original GHOST paper:
\begin{itemize}
    \item The original paper stated, ``Oversampling is effective and necessary prior to applying deep learning for defect prediction.'' We find this is also true for code smell detection
    \item The original paper said, ``We take care to stress that our results relate to defect
prediction. As to other areas of software analytics, that is
a matter for future search.'' This paper is one such future work that extends the prior study by evaluating it on a new domain.
\end{itemize}

\section*{Acknowledgements}

This work was funded by an NSF Award \#1908762.




\balance
\bibliographystyle{ACM-Reference-Format}
\bibliography{sample-base}




\end{document}